# Impact of Oxidation State on the Valence-bond-glass Physics in the lithium-intercalated $Mo_3O_8$ Cluster Mott Insulators

Daigo Ishikita, Yuya Haraguchi[†], and Hiroko Aruga Katori
Department of Applied Physics and Chemical Engineering, Tokyo University of Agriculture and Technology, Koganei, Tokyo 184-8588, Japan

†Corresponding author: chiyuya3@go.tuat.ac.jp

We have successfully synthesized four $Mo_3O_8$-type cluster Mott insulators (CMI) by intercalating lithium into nonmagnetic precursors to regulate the $Mo_3$ cluster valence. The resulting materials are $Li_{1+x}RMo_3O_8$ ($R$ = Sc, Y, Lu) and $Li_xZn_2Mo_3O_8$. Our magnetic susceptibility measurements revealed that these materials display characteristics akin to a valence bond glass state and suggest the presence of short-range ordering when the $Mo_3$ cluster valence approximates its ideal value. These findings challenge the prevailing belief that the plaquette charge ordering state is an inherent feature of $Mo_3O_8$-type CMI. Instead, they underscore the importance of $Mo_3$ cluster valence in determining the physical properties of these systems. These insights furnish a fresh understanding of the $Mo_3O_8$-type CMI and open new research opportunities in highly frustrated magnetism.

## I. Introduction

Quantum spin-liquids (QSLs) have garnered significant attention in condensed matter physics due to their potential to host exotic quantum phenomena such as fractionalized excitations and long-range entanglement [1-4]. These QSL materials, characterized by highly frustrated spin configurations and the absence of conventional long-range magnetic order, have posed a challenge for both experimentalists and theoreticians, demanding the development of innovative synthetic methods and advanced observation tools for QSLs. Moreover, the breakthrough of these material discoveries offers distinct perspectives on quantum physics [5-9].

$Mo_3O_8$-type cluster Mott insulators (CMIs) are promising candidates for realizing QSLs. These compounds are comprised of Mo atoms arranged in kagome layers trimerized to form a triangular lattice of $Mo_3$ clusters, as shown in Fig. 1. Studies of $LiZn_2Mo_3O_8$ have revealed an intriguing phenomenon: the Curie constant at low temperatures is observed to be one-third of its value at high temperatures, implying a diminish of two-thirds of the total spins [10]. An initial explanation for this partial spin disappearance mechanism was proffered by forming a spin-singlet on the triangular lattice of a honeycomb superlattice, with the remaining one-third of the spin constituting a valence bond solid (VBS) [11-13]. Subsequently, the $Mo_3$ CMIs were reinterpreted as the extended Hubbard model on the 1/6 filled breathing kagome lattice, leading to the development of a theoretical framework to describe the electronic state of the system, postulating the plaquette charge order (PCO) state as the ground state [14-16]. In the PCO state, three unpaired electrons are arranged in a hexagonal configuration, with each electron placed at a vertex and not adjacent to its neighbors. However, the effect of the PCO state on the magnetic order is not yet understood. While it accounts for the reduced spin in $LiZn_2Mo_3O_8$, its role in the emergence of a quantum spin liquid state is less clear, given the frustrated geometry of the lattice. The potential for non-nearest neighbor interactions to suppress magnetic order invites further scrutiny. The connection between the PCO state and the suppression of magnetic order in $Mo_3O_8$-type CMIs is still under investigation. The presumption that spins on a triangular lattice naturally form a spin liquid state is questionable without factoring in extended interactions. Thus, future research explorings, exemplified by Flint and Lee's emergent honeycomb lattice model [11], are essential to discern if diminished super-exchange interactions are solely responsible for inducing a spin liquid state or if other dynamics contribute.

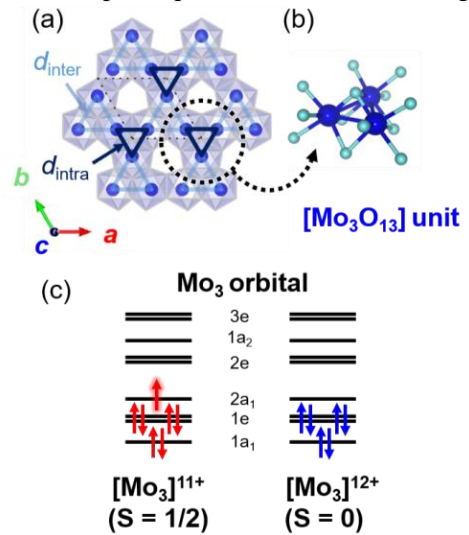

**Fig. 1** Triangular lattice constituted by $Mo_3$ clusters, accompanied by their respective molecular orbitals and electron configurations corresponding to their valence states.

Alternatively, recent structural investigations grounded in solid-state chemistry have revealed significant off-stoichiometry in $LiZn_2Mo_3O_8$ [17,18]. The magnetic susceptibility of $LiZn_2Mo_3O_8$ with nearly perfect stoichiometric ratios is not reproduced at all by the PCO model. Instead, it displays magnetic susceptibility characteristics consistent with strongly correlated two-dimensional antiferromagnetic systems [18]. The diminution of the PCO-like behavior as the stoichiometry improves raises questions regarding the validity of the previous hypothesis. While other $Mo_3O_8$-type CMIs like $Li_2InMo_3O_8$ [19] demonstrate magnetic ordering, the characterization of $Li_2ScMo_3O_8$ [19] is more complex. Although it was initially thought to exhibit QSL-like behavior probed in the NMR study, subsequent μSR experiments have provided clear evidence of spin freezing [15,20]. Similarly, for $Na_3A_2(MoO_4)_2Mo_3O_8$ (A = In, Sc) [21,22], the observed magnetic properties cannot be adequately explained by the PCO model alone, indicating the need for a revised understanding or additional theoretical models to account for these behaviors.

As mentioned, despite a spin defect associated with offstoichiometry in $Mo_3O_8$-type CMIs, the experimental results are consistent with theoretical predictions and currently represent the physical properties of $Mo_3O_8$-type CMIs. As shown in Fig. 1(c), magnetic/nonmagnetic characteristics in a $Mo_3$ cluster are highly susceptible to the valence of molybdenum ion. However, the effect of spin defects caused by deviations from the ideal valence of $Mo_3$ clusters has not been thoroughly examined. Consequently, we aimed to shed light on the fundamental properties of $Mo_3O_8$-type CMIs by synthesizing novel materials through precise control of the valence of $Mo_3$ clusters, which would be a crucial determinant of the physical properties of these systems.

This paper reports the successful synthesis of four $Mo_3O_8$-type CMIs using Li-intercalation to control $Mo_3$ cluster valence. Our magnetic susceptibility results showed valence bond glass-like behaviors and short-range ordering as $Mo_3$ cluster valence approached ideal value, contradicting previous understanding of PCO state as an indispensable attribute of $Mo_3O_8$-type CMIs and highlighting $Mo_3$ cluster valence as a crucial determinant of these systems' properties. Our findings offer new insight into $Mo_3O_8$-type CMIs and the potential for further advancement in frustrated physics.

**TABLE 1** Comparison of structural parameters for precursor and Li-intercalated samples.

|  | Space group | $a$ [Å] | $d$ [Å] |
|---|---|---|---|
| $LiScMo_3O_8$ | $P3m1$ | 5.724(1) | 4.943(1) |
| $3R$-$Li_{1+x}ScMo_3O_8$ | $R3m$ | 5.77193(7) | 5.137 59(3) |
| $LiLuMo_3O_8$ | $P3m1$ | 5.766(1) | 5.096(1) |
| $Li_{1+x}LuMo_3O_8$ | $P3m1$ | 5.7673(5) | 5.207 43(5) |
| $LiYMo_3O_8$ | $P3m1$ | 5.781(1) | 5.153(2) |
| $Li_{1+x}YMo_3O_8$ | $P3m1$ | 5.767 43(9) | 5.260 56(10) |
| $Zn_2Mo_3O_8$ | $P6_3mc$ | 5.7835(2) | 4.9498(3) |
| $2H$-$Li_xZn_2Mo_3O_8$ | $P6_3mc$ | 5.731 81(2) | 5.168 28(5) |

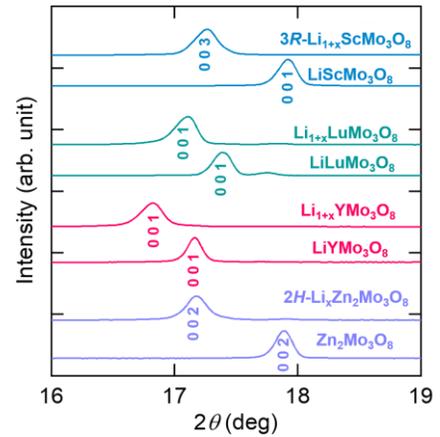

**Fig. 2** X-ray diffraction patterns of $LiRMo_3O_8$ ($R$ = Sc, Lu, Y) and $Zn_2Mo_3O_8$ and their lithiated form focused on the angular range of 16 < 2θ/deg < 19. Intensities have been normalized. Peaks with Miller indices indicate the interlayer spacings $d$.

## II. Experimental Methods

The precursors $LiRMo_3O_8$ ($R$ = Sc, Y, Lu) and $Zn_2Mo_3O_8$ were synthesized by conventional solid-state reaction. $Li_2RMo_3O_8$ and $LiZn_2Mo_3O_8$ were synthesized by Li intercalation of the obtained precursors according to the following chemical reaction,

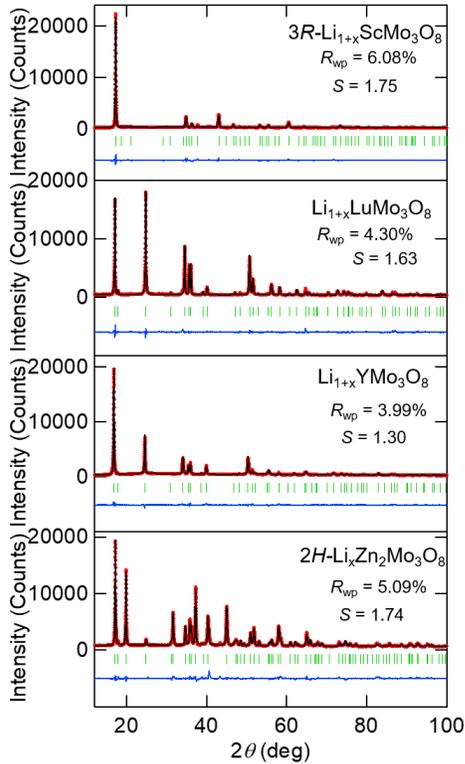

**Fig. 3** X-ray diffraction patterns of lithiated Li$R$Mo$_3$O$_8$ ($R$ = Sc, Lu, Y) and Zn$_2$Mo$_3$O$_8$. The green vertical lines indicate the Bragg reflection positions of the intercalated samples. The observed intensities (red circles), calculated intensities (black line), and their differences (blue curve at the bottom) are shown.

$$\text{Li}R\text{Mo}_3\text{O}_8 + \text{LiH} \rightarrow \text{Li}_2R\text{Mo}_3\text{O}_8 + 0.5\text{H}_2.$$
$$\text{Zn}_2\text{Mo}_3\text{O}_8 + \text{LiH} \rightarrow \text{LiZn}_2\text{Mo}_3\text{O}_8 + 0.5\text{H}_2.$$

The precursors and four-molar excess of LiH were finely ground in an Ar-filled glove box, sealed in an evacuated Pyrex tube, and heated at 300$^\circ$C. The optimal molar ratio for the LiH mixture was determined through experimentation. When the molar amount of LiH was approximately one to three times the stoichiometric ratio, the reaction was incomplete with a remaining unreacted precursor. On the other hand, when the LiH amount was large excess (about tenfold the stoichiometric ratio), the reduction reaction proceeded further, resulting in the deposition of the molybdenum element. After the reaction, the pellets were washed with methanol to remove residual LiH. The thus-obtained sample was characterized by powder X-ray diffraction (XRD) experiments in a diffractometer with Cu-K$\alpha$ radiation. The cell parameters and crystal structure were refined by the Rietveld method using the Z-Rietveld v1.1.3 software [24,25] (see Supplemental Material [26]). The temperature dependence of magnetization was measured under several magnetic fields in a magnetic property measurement system.

### III. Results

#### A. Structural analysis

Using X-ray diffraction analysis, as illustrated in Fig. 2, we observed shifts in the lowest 00$l$ peaks associated with the Mo$_3$O$_8$ interlayer spacing after Li-intercalation. In addition, the shifts moved toward lower angles, as shown by comparing the lattice constants of these materials. This change suggests that the intercalation of lithium ions into the samples resulted in a noticeable elongation of interlayer spacing. The observed structural modifications from the precursor materials to lithium intercalated them are summarized in Table 1. A common change across all samples was the noticeable increase in the interlayer space $d$ after the lithium insertion process. This expansion in interlayer space provides strong evidence that lithium ions have indeed been successfully inserted into the interlayer regions of these materials.

To comprehensively investigate the structural changes caused by lithium insertion, we performed Rietveld analysis on all samples, as depicted in Fig. 3. The detailed structural parameters are listed in the Supplementary Materials. The results revealed that Li$_{1+x}$LuMo$_3$O$_8$, Li$_{1+x}$YMo$_3$O$_8$, and 2$H$-Li$_x$Zn$_2$Mo$_3$O$_8$ retained the same space group as their precursors ($x$ = 0), indicating no change in their crystal structure. On the contrary, for 3$R$-Li$_{1+x}$ScMo$_3$O$_8$, there was a shift in the space group from $P3m$1 to $R3m$, as detected in the x-ray diffraction patterns. It is important to note that the space groups of 3$R$-Li$_2$ScMo$_3$O$_8$ and 2$H$-LiZn$_2$Mo$_3$O$_8$ differed from those synthesized via the high-temperature ceramics method.

Figure 4 depicts the crystal structure transformations through lithium intercalation. LiYMo$_3$O$_8$, LiLuMo$_3$O$_8$, and Zn$_2$Mo$_3$O$_8$ retained their crystal structures, showing identical space groups before and after lithiation. On the other hand, in the case of LiScMo$_3$O$_8$, we noticed structural alterations post-lithiation. The precursor LiScMo$_3$O$_8$ presents a $P3m$1 space group structure akin to LiYMo$_3$O$_8$ and LiLuMo$_3$O$_8$. However, the layer stacking pattern changed to an $R3m$ structure upon lithium insertion. This observation suggests that the relatively lighter Sc$^{3+}$ ions likely migrate within the crystal structure due to the lithium insertion process, resulting in this remarkable structural transformation.

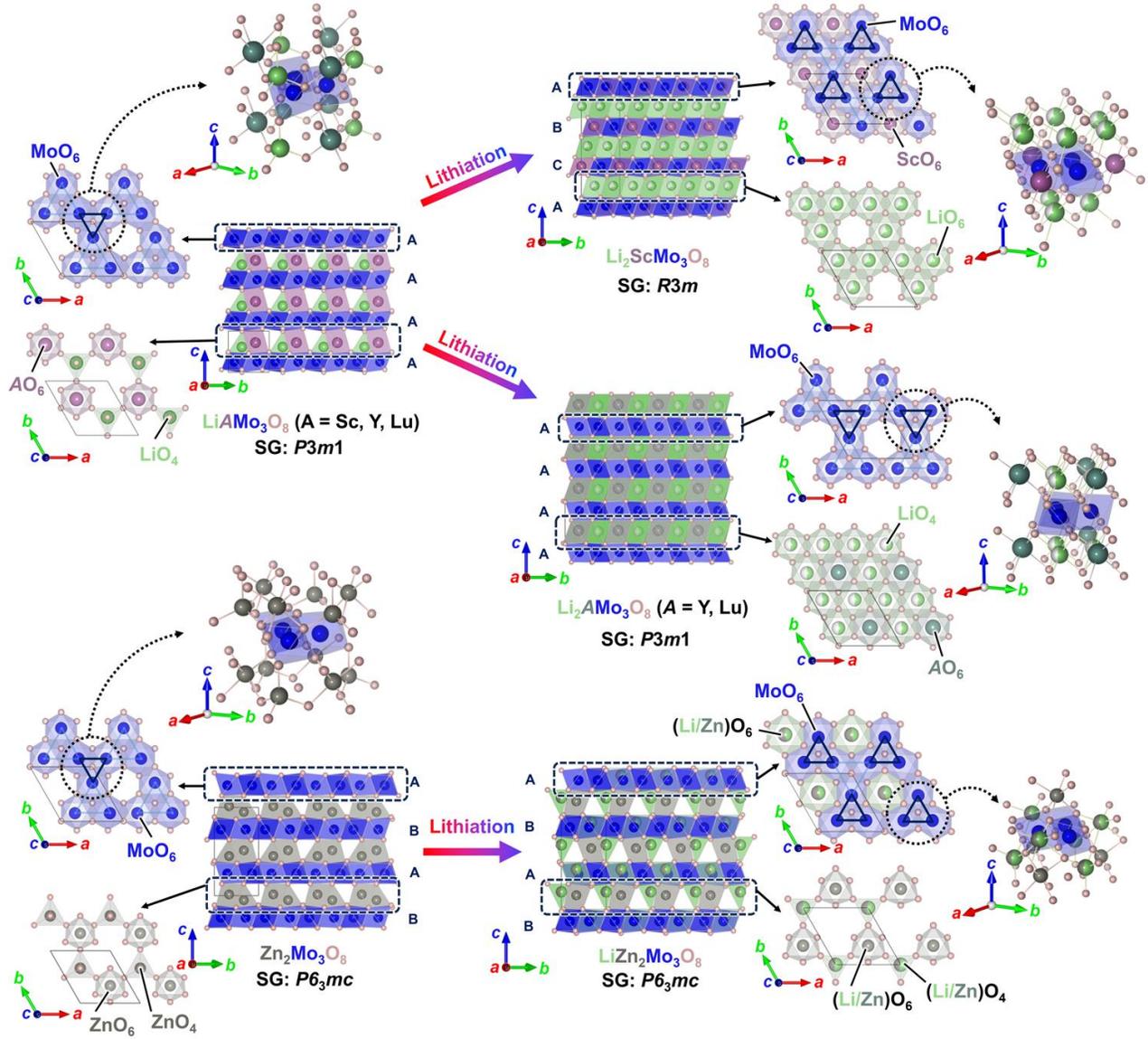

**Fig. 4** The crystal structure of nonmagnetic precursors Li$R$Mo$_3$O$_8$ ($R$ = Lu, Y, Sc) and $A_2$Mo$_3$O$_8$ ($A$= Mg, Zn) and their lithiated products Li$_2R$Mo$_3$O$_8$ ($R$ = Lu, Y, Sc) and Li$A_2$Mo$_3$O$_8$ ($A$= Mg, Zn) along with the transformation. The layer-by-layer crystal structure and the local environment around the Mo$_3$O$_{13}$ cluster of each compound are also shown. The VESTA program is used for visualization [23].

**B. Magnetization measurement**

The temperature dependences of the magnetic susceptibility $M/H$ and the inverse magnetic susceptibility $H/M$ of the intercalated samples are depicted in Fig. 5, with the measurement results of the precursors included for comparison. The measurements were conducted under an applied external magnetic field of 1 T. The linearity observed in the high-temperature region of the inverse susceptibility suggests the presence of localized spins.

We performed Curie-Weiss fitting to $\chi$-data at high temperatures (200-300 K), yielding the effective magnetic moments ($\mu_{\text{eff}}^{\text{HT}}$) and Weiss temperatures ($\theta_{\text{W}}^{\text{HT}}$), as summarized in TABLE 2. All intercalated samples exhibited an enhancement in the $\mu_{\text{eff}}^{\text{HT}}$-value compared to their precursors. This enhancement must be linked to the Li intercalation, which is theorized to cause a reduction in the valence of the Mo$_3$ cluster from 12+ ($S = 0$) to 11+ ($S = 1/2$). The effective magnetic moment for the precursors is significantly smaller than Li-intercalated samples but not the expected value of zero. This unexpected finding suggests that there is inherent offstoichiometry also in the precursor and that the Mo clusters are not in their ideal valence states, with evidence of slight valence change from ideal [Mo$_3$]$^{12+}$. Surprisingly, there appears to be a substantial magnetic correlation among the diluted spins in the Mo$_3$ clusters of the precursor, as probed by the significantly large negative Weiss temperature. This observation is distinct from the Curie-tail behavior characterized by an almost zero Weiss temperature at low temperatures, as seen in the Li-intercalated compounds discussed subsequently. Such a large Weiss temperature has also been observed in Li$_2$Sc$_{1-x}$Sn$_x$Mo$_3$O$_8$, which exhibits significant spin dilution [27].

**TABLE 2** The Mo$_3$ cluster valences, high-temperature and low-temperature $\mu_{eff}$, and $\theta_W$ values for lithiated products.

| sample | $x =$ | [Mo$_3$]$^{n+}$ | $\mu_{eff}^{HT}$/Mo$_3$($\mu_B$) | $\theta_W^{HT}$ (K) | $\mu_{eff}^{LT}$/Mo$_3$($\mu_B$) | $\theta_W^{LT}$ (K) | reference |
|---|---|---|---|---|---|---|---|
| ideal | - | 11 | $\sqrt{3}$ | - | 1 | - | - |
| 6R-Li$_{0.95}$Zn$_{1.92}$Mo$_3$O$_8$ | - | 11.11 | 1.63(12) | −440(110) | 0.607(2) | −5.2 | [18] |
| 3R-Li$_{1+x}$ScMo$_3$O$_8$ | 0.86 | 11.14 | 1.609(5) | −215(5) | 0.552(8) | −5.6(5) | this work |
| 2H-Li$_x$Zn$_2$Mo$_3$O$_8$ | 0.77 | 11.23 | 1.521(2) | −202(2) | 0.338(7) | −3.9(3) | this work |
| Li$_{1+x}$LuMo$_3$O$_8$ | 0.77 | 11.23 | 1.516(2) | −103.7(9) | 0.732(8) | −3.8(2) | this work |
| 6R-Li$_{1.2}$Zn$_{1.8}$Mo$_3$O$_8$ | - | 11.36 | 1.39 | −220 | 0.8 | −14 | [10] |
| Li$_{1+x}$YMo$_3$O$_8$ | 0.62 | 11.38 | 1.362(7) | −211(5) | 0.603(2) | −0.78(7) | this work |

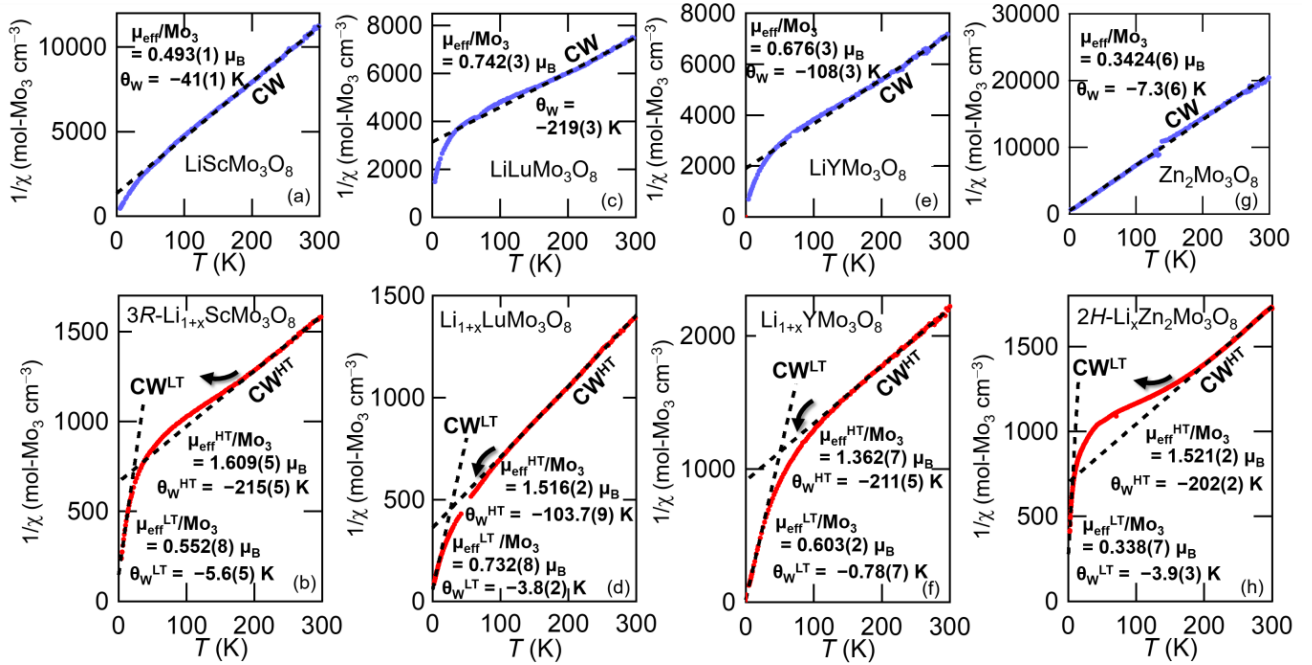

**Fig. 5** Temperature dependence of inversed magnetic susceptibility $M/H$ for the following compounds: (a) LiScMo$_3$O$_8$, (b) 3R-Li$_{1+x}$ScMo$_3$O$_8$, (c) LiLuMo$_3$O$_8$, (d) Li$_{1+x}$LuMo$_3$O$_8$, (e) LiYMo$_3$O$_8$, (f) Li$_{1+x}$YMo$_3$O$_8$, (g) Zn$_2$Mo$_3$O$_8$, and (h) 2H-Li$_x$Zn$_2$Mo$_3$O$_8$. The dotted lines in all panels indicate the results of Curie-Weiss fits. The sinuous arrows delineate the temperature regions wherein the observed $H/M$ curves exhibit deviation from the Curie-Weiss fit, distinguishing the manner of deviation by convex up and down arrows.

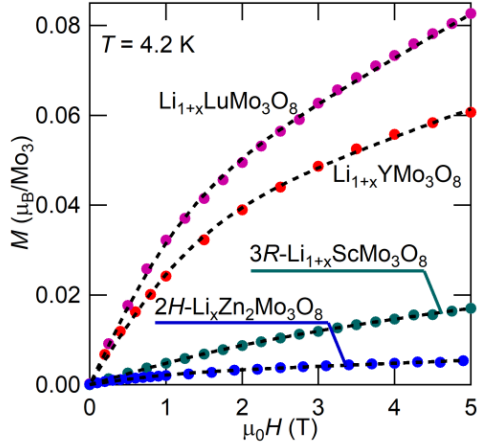

**Fig. 6** The isothermal magnetization of at 4.2 K (circles) and those fitting curves (see the text).

TABLE 2 presents the inferred Mo$_3$ cluster valence and the projected chemical formula based on the effective magnetic moment values. The Mo$_3$ cluster valence was calculated by substituting the μ$_{eff}$ value into the equation as follows,

$$n = 12 - \left(\frac{\mu_{eff}^{HT}}{2\sqrt{S(S+1)}}\right)^2 = 12 - \left(\frac{\mu_{eff}^{HT}}{\sqrt{3}}\right)^2. \qquad (1)$$

Here, the methodology employed was to iteratively compute the μ$_{eff}$ value from the assumed chemical formula and subsequently determine the chemical formula corresponding to the calculated μ$_{eff}$-value until convergence (details are provided in the Supplementary Materials [26]). As a result, the calculated μ$_{eff}$-value of all samples is less than the expected ~1.73 μ$_B$ for $S = 1/2$, indicating an incomplete occupancy of lithium ions at the vacancy sites of precursors.

As shown in Fig. 5, there is an evident deviation from the Curie-Weiss (CW) law at low temperatures, which is particularly noticeable in the behavior of the inverse susceptibility as the temperature decreases. However, as the temperature drops, a trend of linear inverse susceptibility with temperature change re-emerges. These behaviors are observed across all materials under study. The CW fitting in the low-temperature region was carried out separately from the fitting in the high-temperature range. The resultant values for the effective magnetic moments (μ$_{eff}^{LT}$) and the Weiss temperatures (θ$_W^{LT}$) obtained from this low-temperature CW fitting are detailed in TABLE 2. The absolute values of θ$_W^{LT}$- and μ$_{eff}^{LT}$ are smaller than θ$_W^{HT}$ and μ$_{eff}^{HT}$. This trend is consistent across all intercalated samples. This phenomenon can be attributed to the partial suppression of magnetic moments commonly observed in LiZn$_2$Mo$_3$O$_8$ [18], Li$_2$Sc$_{1-x}$In$_x$Mo$_3$O$_8$ [15], and Li$_2$Sc$_{1-x}$Sn$_x$Mo$_3$O$_8$ [27].

The behavior of the inverse susceptibility in the intermediate temperature range displays deviations from the CW line in all Li-intercalated Mo$_3$O$_8$ materials. These deviations, however, are not uniform across the different materials and can be grouped into two distinct classes. The first group, Li$_{1+x}$LuMo$_3$O$_8$ and Li$_{1+x}$YMo$_3$O$_8$, displays a downward deviation from the CW line as temperature decreases. Conversely, the second group, comprising 3R-Li$_{1+x}$ScMo$_3$O$_8$ and 2H-Li$_x$Zn$_2$Mo$_3$O$_8$, shows an upward deviation from the CW line. Differences in behavior across various materials can be linked to their respective μ$_{eff}$-values—The first group materials exhibit relatively small μ$_{eff}$ values, whereas the second group demonstrates larger μ$_{eff}$ values. The intricate relationship between these values and the observed susceptibility behaviors will be discussed in subsequent sections.

Figure 6 shows the isothermal magnetization measured at $T = 4.2$ K. All samples exhibit magnetization curves characterized by upward convexity, indicative of free spin presence. To quantify the free spin amount, we employed a magnetization model for analysis comprising two components: a quasi-free spin component modeled by the Brillouin function of $S = 1/2$ and a strongly correlated spin component, which is linearly dependent on the applied magnetic field.

$$M = \chi H + fN_A\mu_B \times B_s[H\mu_B/k_B(T - \theta_w)], \qquad (2)$$

where $\chi$ corresponds to the magnetic susceptibility attributable to correlated spins, and $\theta_W$ represent Weiss temperature associated with the nearly free spins. The symbols $N_A$, $k_B$, and $B_s$ denote Avogadro number, Boltzmann's constant, and Brillouin function, respectively. The parameter $f$ quantifies the proportion of free spins. The optimal fits are illustrated by the dashed lines in Fig 6, with the corresponding fitting parameters presented in TABLE 3. The ability of this model to accurately characterize the magnetization process suggests the presence of free spins at low temperatures, which is also probed by nearly zero Weiss temperature. However, the data do not provide enough information to ascertain whether these spins are

**TABLE 3** Results of fits to the isothermal magnetization data using the formula described in the text. The fitting parameters $g$ and $S$ are fixed to 2 and 0.5, for respectively.

| Sample | $f$(%) | θ$_W$(K) | χ(cm$^3$/mol-Mo$_3$) |
|---|---|---|---|
| 3R-Li$_{1+x}$ScMo$_3$O$_8$ | 0.61(6) | 2.68(1) | 1.23(8)×10$^{-3}$ |
| Li$_{1+x}$LuMo$_3$O$_8$ | 3.4(1) | 3.31(4) | 5.4(2)×10$^{-3}$ |
| Li$_{1+x}$YMo$_3$O$_8$ | 3.1(2) | 3.22(7) | 3.4(5)×10$^{-3}$ |
| 2H-Li$_x$Zn$_2$Mo$_3$O$_8$ | 0.21(2) | 3.45(8) | 3.8(5)×10$^{-4}$ |

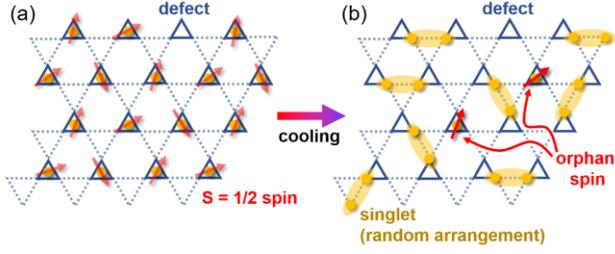

**Fig. 7** (a) The partial spin defect model on a triangular lattice constituted by Mo$_3$ clusters. The Mo$_3$ clusters are depicted as equilateral triangles. On these triangles, yellow circles with red arrows are placed, symbolizing unpaired electrons corresponding to the S = 1/2 spin derived from the [Mo$_3$]$^{11+}$ cluster. The unmarked triangle represents the [Mo$_3$]$^{12+}$ cluster referred to as a nonmagnetic defect due to its lack of unpaired electrons. (b) The formation diagram of VBG. Yellow ellipses signify spin-singlet pairs, indicating paired electrons with opposite spins that result in no net magnetic moment. The isolated spins, on the other hand, are still represented by red arrows.

extrinsic, potentially due to defects or impurities, or if they are an intrinsic feature of the system. Thus, the observed Curie-tail-like divergence in the temperature dependence of magnetic susceptibility can be attributed to the presence of nealy free spins.

## IV. Discussion

The estimated $\mu_{eff}^{HT}$ value in all lithiated Mo$_3$O$_8$-type CMIs was less than anticipated for the $S = 1/2$ spin state. As displayed in TABLE 2, this discrepancy suggests that some Mo$_3$ clusters possess their nonmagnetic [Mo$_3$]$^{12+}$ state rather than magnetic [Mo$_3$]$^{11+}$. This situation is visually represented using an $S = 1/2$ dilute triangular lattice model, shown in Fig. 7(a). The magnetism of CMI in the high-temperature region is consistent with the CW law with strong antiferromagnetic interactions. Interestingly, CMIs stray from the Curie-Weiss law in the intermediate temperature range as the temperature decreases, only to return to compliance in the low-temperature region. The difference between the two CW regimes is marked by a decline in the $\mu_{eff}$ and $\theta_W$ values from high- to low temperatures.

The valence bond glass (VBG) model reasonably explains the observed behaviors [28, 29]. This model suggests that influences such as spin defects can instigate bond randomness. This randomness creates spin singlets in a disordered state, leaving isolated spins at overlooked sites during singlet formation, as visualized in Fig. 7(b). This occurrence helps interpret the noticed reduction in spin behaviors as temperature decreases. At high temperatures, the system shows clear paramagnetism. However, as temperatures decrease, a deviation from the CW law is noted, correlating with the gradual formation of spin singlets due to strong spin interactions. At low temperatures, isolated spins manifest in a VBG state. These states align with the two CW regimes and offer a plausible explanation for the decrease in both $\mu_{eff}$ and $|\theta_W|$ values observed during our experiments. Remarkably, as shown in Fig. 5, the $H/M$ curves of the precursors also display a two CW regime, which persists even with a low concentration of spins. This phenomenon could potentially be accounted for by the VBG mechanism.

LiZn$_2$Mo$_3$O$_8$ also exhibits a spin reduction behavior, as previously documented [10]. The PCO mechanism has been proposed to elucidate the magnetism of LiZn$_2$Mo$_3$O$_8$ [14]. In the PCO state, three unpaired electrons, referred to as "plaquettes," are distributed evenly across the lattice, ensuring that no two plaquettes share a vertex of the hexagon formed by three Mo$_3$ clusters. This model successfully explains the magnetism that appears to be a 2/3 spin reduction. The magnetic properties of LiZn$_2$Mo$_3$O$_8$, particularly the high-temperature $\mu_{eff}$ and $\theta_W$ values, are similar to those of Li$_{1+x}$YMo$_3$O$_8$ and Li$_{1+x}$LuMo$_3$O$_8$. The present study has successfully applied the VBG model to describe the magnetic properties of Li$_{1+x}$YMo$_3$O$_8$ and Li$_{1+x}$LuMo$_3$O$_8$, suggesting the presence of partial spin defects in these compounds. This scenario is further supported by the temperature dependence of inverse susceptibility in α-LiZn$_2$Mo$_3$O$_8$ [10], which aligns qualitatively and quantitatively with the magnetism observed in Li$_{1+x}$Ymo$_3$O$_8$ and Li$_{1+x}$LuMo$_3$O$_8$. Notably, subsequent investigations have revealed a significant Li/Zn off-stoichiometry in LiZn$_2$Mo$_3$O$_8$ [18].

Here, we leave aside the possibility of VBG and consider whether the PCO model can explain each. The temperature dependence of the magnetic susceptibility in the PCO state is predicted by the following equation,

$$\chi_{PCO} = \frac{\beta(T)C}{T - \beta(T)\theta_W} \left( \beta(T) = \frac{1 + 5\exp(-\Delta E/k_B T)}{1 + \exp(-\Delta E/k_B T)} \right) \quad (3)$$

, where $C$ is the Curie constant, and $\Delta E$ is the energy gap between the ground state with $S_{tot} = 1/2$ and the excited state with $S_{tot} = 3/2$, comprised of three spins forming a plaquette. The "phenomenological PCO model" derived by Akbari-Sharbaf et al. [15] significantly simplifies the theoretical ideas originally proposed by Chen et al. [14], yet it is known to reproduce experimental data well. We fit the $H/M$ curves of Li-intercalated Mo$_3$O$_8$ compounds to the phenomenological PCO model with equation (3), as shown in Fig. 8. The phenomenological PCO model provides a satisfactory approximation of the magnetic susceptibility only for $Li_{1+x}YMo_3O_8$ and $Li_{1+x}LuMo_3O_8$. However, it fails to accurately represent the $\chi$-data for $3R$-$Li_{1+x}ScMo_3O_8$ and $2H$-$Li_xZn_2Mo_3O_8$. This discrepancy suggests that these materials manifest magnetic states beyond the predictive capability of the phenomenological PCO model. Furthermore, while the effective magnetic moments estimated by the phenomenological PCO fit for $Li_{1+x}YMo_3O_8$ approximately agree with that by the CW fit, this is not true for $Li_{1+x}LuMo_3O_8$, which is seemingly well fit by the phenomenological PCO model. This variation undermines the reliability of the phenomenological PCO fit.

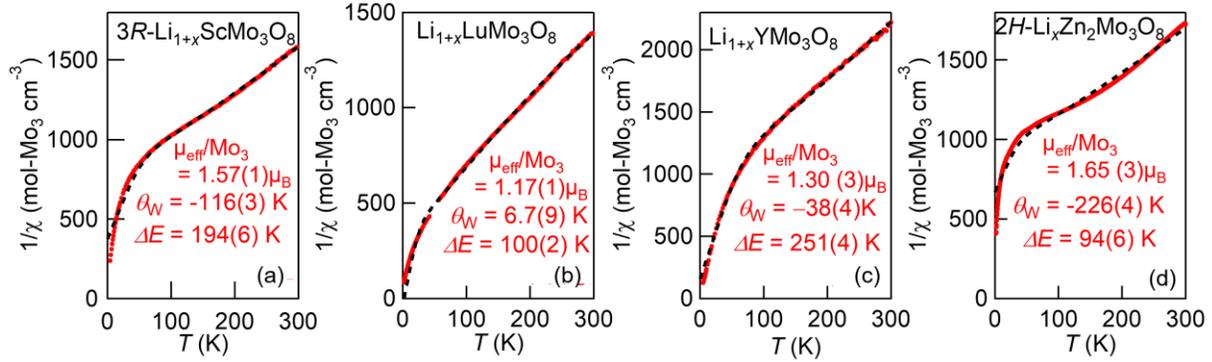

**Fig. 8** Results obtained from fitting the phenomenological PCO model to the $H/M$ data for (a) $Li_{1+x}ScMo_3O_8$, (b) $Li_{1+x}LuMo_3O_8$, (c) $Li_{1+x}YMo_3O_8$, (d) $Li_xZn_2Mo_3O_8$. Values near the fitting result curves are fitting parameters (see the main text).

The reproducibility of the phenomenological PCO model for $Mo_3O_8$-type CMIs is limited, as its applicability is restricted to certain materials such as $Li_{1+x}YMo_3O_8$ with a $\mu_{eff}^{HT}$-value of approximately 1.4 $\mu_B$. This constraint also applies to $6R$-$LiZn_2Mo_3O_8$ [10] and $Li_2Sc_{0.6}In_{0.4}Mo_3O_8$ [15], each exhibiting $\mu_{eff}^{HT}$ values around 1.4 $\mu_B$. In addition, The phenomenological PCO model accurately replicates the entire temperature profile of the inverse susceptibility for $Li_2Sc_{0.8}Sn_{0.2}Mo_3O_8$ ($\mu_{eff}^{HT}$ = 1.39 $\mu_B$), in which the valence is deliberately decreased by substituting $Sc^{3+}$ with $Sn^{4+}$ of $2H$-$Li_2ScMo_3O_8$, as detailed in the Supplementary Materials [26]. On the other hand, the phenomenological PCO model cannot accurately represent the magnetization of three particular compounds, $3R$-$Li_{1+x}ScMo_3O_8$ and $2H$-$Li_xZn_2Mo_3O_8$. This discrepancy is primarily due to the appearance of antiferromagnetic short-range order (SRO), a behavior not accounted for in the phenomenological PCO model. Similar observations have been noted in the optimally synthesized α-$LiZn_2Mo_3O_8$ sample with an almost perfect $\mu_{eff}^{HT}$ value of 1.73, demonstrating SRO without PCO behavior, as reported by Sandvik et al. [18]. Consequently, these findings bring into question the validity of the phenomenological PCO model, especially for the $[Mo_3]^{11+}$ state, considered ideal within this framework. Furthermore, these results suggest that attaining the predictions of the PCO mechanism might be impracticable in situations devoid of spin defects.

Our study reveals several key findings about $Mo_3O_8$-type CMIs. Firstly, our findings indicate that magnetic susceptibility curves consistent with the phenomenological PCO model are detectable within a limited $Mo_3$ valence range characterized by notable spin defects.

Secondly, our study showed a magnetization divergence in lower temperature regions, likely due to VBG-like states arising from disorder. Furthermore, a previous exact diagonalization-based theoretical investigation reported by Watanabe has revealed a notable coexistence of short-range order and emergent free spin in a spin 1/2 random Heisenberg antiferromagnetic model on a triangular lattice [29], which aligns with our observations.

Thirdly, we observed that an SRO emerges as the $Mo_3$-cluster valence nears its ideal $[Mo_3]^{11+}$ state, suggesting the gradual realization of an $S = 1/2$ Heisenberg triangular lattice antiferromagnetic model. However, defects have significant implications: although the system can support the emergence of SRO with defects, these defects can also foster the development of VBG-like states, triggering random dimer formation.

As mentioned, the magnetic characteristics of $Mo_3O_8$-type CMIs are significantly influenced by valence. Figure 9 encapsulates that two-dimensional short-range order is discernible in systems where the effective magnetic moment is close to $\mu_{eff}$ = 1.73 $\mu_B$ expected for $S = 1/2$ spins. Conversely, systems including highly concentrated spin defects with $\mu_{eff}$ ranging from 1.3 to 1.4 $\mu_B$ show the $\chi$-curves as seemingly characterized for the PCO state. Hence, depending on the valence within the $Mo_3$ clusters, the magnetic behavior can be bifurcated into two distinct categories. Contrary to theoretical predictions for $Mo_3O_8$-type CMIs, which posit that magnetism is predominantly influenced by the breathing parameter λ (defined as the ratio of inter-cluster Mo-Mo $d_{inter}$ to intra-cluster Mo-Mo bond distances $d_{intra}$), our observations, in conjunction with previous reports, suggest that λ does not systematically correlate with variations of magnetic ground states. However, λ would be associated with the valence of the $Mo_3$ clusters. As shown in Fig. 1(c), a valence state of $[Mo_3]^{11+}$ ($S = 1/2$) results in a single unpaired electron in the bonding orbital $2a_1$, whereas for the nonmagnetic $[Mo_3]^{12+}$ ($S = 0$) state, this orbital is unoccupied. The difference in valence modifies the "bond order" within the $Mo_3$ clusters; a higher bond order diminishes the Mo-Mo bond length to lower the energy level of the $2a_1$ orbital. Then, an elevation in the bond order would concomitantly increase the λ parameter. Consequently, the λ parameter varies solely with the valence of the $Mo_3$ cluster and appears not to serve as a direct determinant of magnetism. It is plausible to assert that magnetism is exclusively regulated by the valence, with the λ parameter being determined solely by the interplay between the valence and the lattice constant.

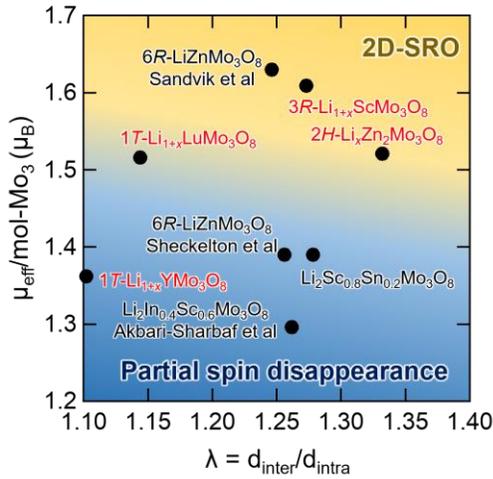

**Fig. 9** Relation between effective magnetic moment, breathing parameter λ and magnetic states in $Mo_3O_8$-type CMIs.

An integrated analysis of our findings supports that the VBG model is a good solution for magnetism in CMIs, although there is no smoking gun to rule out the PCO model. Therefore, the simplicity and directness of the VBG model make it a compelling approach to understanding the mysterious magnetism in CMIs.

## V. Summary

We successfully synthesized five new $Mo_3O_8$-type CMIs: $Li_{1+x}RMo_3O_8$ ($R$ = Sc, Y, or Lu) and $Li_xZn_2Mo_3O_8$, achieved by lithiation into nonmagnetic precursors, thus controlling the valence of the $Mo_3$ clusters in the process. Following synthesis, we then performed magnetization measurements on these $Mo_3O_8$-type CMIs. The results revealed some intriguing properties: when the $Mo_3$ cluster valence neared the ideal value, the materials exhibited both VBG behaviors and evidence of short-range ordering. These findings are notable because they question previous beliefs about the essential properties of $Mo_3O_8$-type CMIs, including the assumed ubiquity of the PCO state. We propose that the valence state of the $Mo_3$ cluster is a critical factor influencing the physical properties of $Mo_3O_8$-type CMIs. Despite the significance of this notion, it has not received much emphasis in prior reports. In conclusion, our study's insights pave the way for further exploration and deepened understanding of frustrated magnetic materials that may contain various chemical defects beyond $Mo_3O_8$-type CMIs.

*Note added.* Recently, we became aware that a similar conceptual study has been reported independently by Wyckoff et al [30]. In this report, $Li^+$ ions were intercalated into $LiScMo_3O_8$ through electrochemical methods, enabling the synthesis of compounds with varying Li contents. Although a detailed interpretation of the temperature dependence of the magnetic susceptibility is not provided, these efforts have been successful in observing both magnetic ordering and changes in the effective magnetic moment that are dependent on the Li composition in compounds near the $Li_2ScMo_3O_8$ composition, with nearly complete control over the Li content. We believe that $Li_xScMo_3O_8$ systems with such continuously variable defects can also be understood within the framework of the VBG model. Furthermore, the fact that $Li_2ScMo_3O_8$ samples synthesized by high-temperature solid-state reaction methods [19] alongside those obtained through the electrochemical method [18] and the current hydride methods exhibit different space groups suggests a polymorphism derivative of the synthesis method. This polymorphism underlines the diverse potential for the further advancement of research in $Mo_3O_8$-type CMIs.

## Acknowledgment

This work was supported by Japan Society for the Promotion of Science (JSPS) KAKENHI Grant Number JP23H04616 (Transformative Research Areas (A) "Supra-ceramics"), JP22K14002 (Young Scientific Research), and JP21K03441 (Scientific Research (C)). Part of this work was carried out by the joint research in the Institute for Solid State Physics, the University of Tokyo.

**Supplementary Information:**
**Impact of Oxidation State on the Valence-bond-glass Physics in Li-Intercalated $Mo_3O_8$ Cluster Mott Insulators**

Daigo Ishikita, Yuya Haraguchi[†], and Hiroko Aruga Katori
Department of Applied Physics and Chemical Engineering, Tokyo University of Agriculture and Technology, Koganei, Tokyo 184-8588, Japan

[†]Corresponding author: chiyuya3@go.tuat.ac.jp

## I. Determination of $\mu_{eff}$ and Chemical Composition

The effective magnetic moment ($\mu_{eff}$) and the chemical composition of a lithium-intercalated sample are determined through the following procedure:

1. We constructed the *H*/*M* data from the raw magnetization measurements, presuming the sample had the ideal chemical composition.
2. We derived the $\mu_{eff}$ value of the sample by applying Curie-Weiss fitting to the *H*/*M* data.
3. We estimated the amount of intercalated lithium using Equation (1) from the main text.
4. We updated the *H*/*M* data incorporating the assumed chemical composition based on the findings from step 3.

This process was iteratively repeated from steps 2 to 4 until both $\mu_{eff}$ and the chemical composition converged, yielding consistent values across three consecutive trials. Figure S1 illustrates the variation in $\mu_{eff}$ values during the iterative Curie-Weiss fitting process, showing convergence at a very early stage. Subsequent analysis reveals negligible correction to the effective magnetic moment throughout this series.

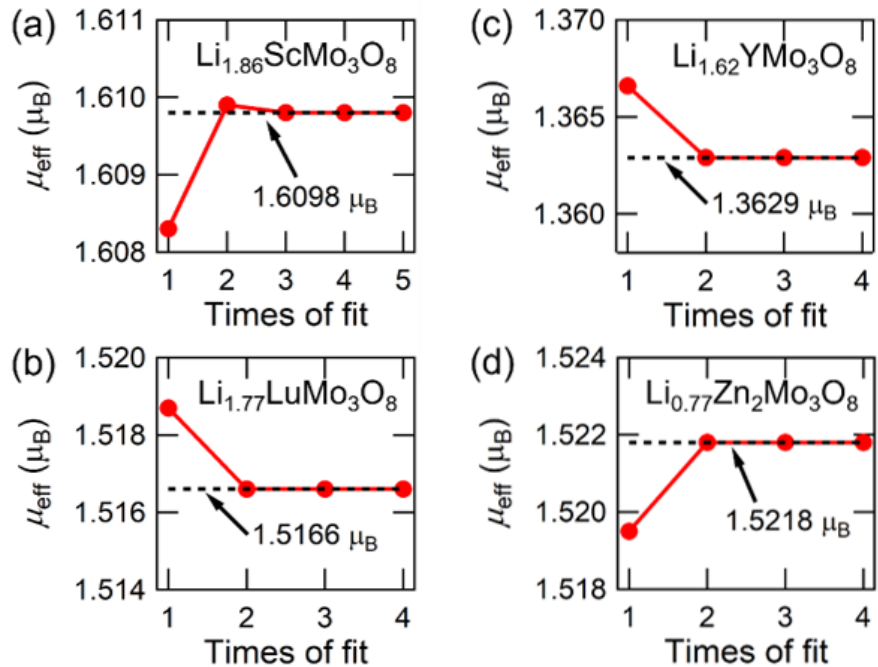

**Figure S1** The final value of $\mu_{eff}$ and chemical composition determined by the iterative Curie-Weiss fitting: (a) $Li_{1+x}ScMo_3O_8$, (b) $Li_{1+x}LuMo_3O_8$, (c) $Li_{1+x}YMo_3O_8$, (d) $Li_xZn_2Mo_3O_8$.

## II. Structure refinement parameters

Table S1 shows the parameters of the Rietveld analysis results shown in Fig. 3 of the main text. The Rietveld refinement estimates the lithium-ion content to be $Li_{1.83}ScMo_3O_8$, $Li_{1.74}LuMo_3O_8$, $Li_{1.58}YMo_3O_8$, and $Li_{0.82}Zn_2Mo_3O_8$, for each compound, respectively, albeit with a considerable margin of error. Nevertheless, this generally agrees with the lithium content estimated from the "self-consistent" magnetic susceptibility analysis, as presented in Figure S1.

**TABLE S1** Crystallographic parameters for $Li_{1+x}RMo_3O_8$ ($R$ = Sc, Lu, Y) and $Li_xZn_2Mo_3O_8$ determined using powder x-ray diffraction. $B$ is the thermal displacement parameter. The obtained lattice parameters are $a$ = 5.771 93(7) Å, 5.7673(5) Å, 5.767 43(9) Å, 5.731 81(2) Å and $c$ = 15.412 78(9) Å, 5.207 43(5) Å, 5.260 56(10) Å, 10.336 56(6) Å, respectively, for $Li_{1+x}ScMo_3O_8$, $Li_{1+x}LuMo_3O_8$, $Li_{1+x}YMo_3O_8$, and $Li_xZn_2Mo_3O_8$.

| | Site | Occ. | $x$ | $y$ | $z$ | $B$ (Å$^2$) |
|---|---|---|---|---|---|---|
| $Li_{1.83}ScMo_3O_8$ | | | | | | |
| Li1 | 9b | 0.61(5) | 0.173(2) | 1-x | 0.1027(16) | 1(fix) |
| Sc1 | 3a | 1 | 0 | 0 | 0.26719(14) | 0.12(4) |
| Mo1 | 9b | 1 | 0.4821(3) | 1-x | 0.26692(13) | 1.23(2) |
| O1 | 9b | 1 | 0.5035(6) | 1-x | 0.00805(12) | 0.275(8) |
| O2 | 9b | 1 | 0.8316(6) | 1-x | 0.18903(11) | 0.77(8) |
| O3 | 3a | 1 | 0 | 0 | 0.5068(3) | 0.41(13) |
| O4 | 3a | 1 | 0 | 0 | -0.0032(4) | 1.47(19) |
| $Li_{1.77}LuMo_3O_8$ | | | | | | |
| Li1 | 3d | 0.58(3) | 0.154(3) | 1-x | 0.237(2) | 1(fix) |
| Lu1 | 1b | 1 | 2/3 | 1/3 | 0.31533(17) | 1.528(15) |
| Mo1 | 3d | 1 | 0.18014(9) | 1-x | 0.79165(18) | 1.413(14) |
| O1 | 1a | 1 | 0.8431(3) | 1-x | 0.5582(8) | 0.50(9) |
| O2 | 1c | 1 | 0.4831(5) | 1-x | 0.0117(7) | 0.40(11) |
| O3 | 3d | 1 | 0 | 0 | -0.0580(11) | 0.38(11) |
| O4 | 3d | 1 | 1/3 | 2/3 | 0.5517(13) | 0.42(17) |
| $Li_{1.62}YMo_3O_8$ | | | | | | |
| Li2 | 3d | 0.56(4) | 0.160(6) | 1-x | 0.292(6) | 1(fix) |
| Y1 | 1b | 1 | 2/3 | 1/3 | 0.3426(2) | 0.33(3) |
| Mo1 | 3d | 1 | 0.17479(16) | 1-x | 0.8064(2) | 0.95(2) |
| O1 | 1a | 1 | 0.832(2) | 1-x | 0.5920(8) | 0.39(11) |
| O2 | 1c | 1 | 0.4851(7) | 1-x | 0.0164(7) | 0.40(15) |
| O3 | 3d | 1 | 0 | 0 | 0.0237(16) | 0.7(3) |
| O4 | 3d | 1 | 1/3 | 2/3 | 0.5573(13) | -0.4(3) |
| $Li_{0.77}Zn_2Mo_3O_8$ | | | | | | |
| Li1 | 2b | 0.40(4) | 1/3 | 2/3 | -0.0547(2) | 0.746(12) |
| Zn1 | 2b | 0.5864(16) | 1/3 | 2/3 | -0.0547(2) | 0.746(12) |
| Li2 | 2a | 0.52(4) | 0 | 0 | 0.2053(3) | 0.71(2) |
| Zn2 | 2a | 0.4136(16) | 0 | 0 | 0.2053(3) | 0.71(2) |
| Zn3 | 2b | 1 | 1/3 | 2/3 | 0.67184(19) | 0.61(3) |
| Mo1 | 6c | 1 | 0.18927(4) | 1-x | 0.39160(15) | 0.955(10) |
| O1 | 2a | 1 | 0.5087(3) | 1-x | 0.0280(7) | 0.55(8) |
| O2 | 2b | 1 | 0.8350(3) | 1-x | 0.2922(4) | 0.44(6) |
| O3 | 6c | 1 | 0 | 0 | 0.0161(10) | 0.70(3) |
| O4 | 6c | 1 | 1/3 | 2/3 | 0.2642(4) | 0.40(14) |

## III. the phenomenological PCO fitting for $Li_2Sc_{0.8}Sn_{0.2}Mo_3O_8$

The phenomenological PCO model fitting, as described by Equation (3) in the main text, was applied to the inverse magnetic susceptibility data for $Li_2Sc_{0.8}Sn_{0.2}Mo_3O_8$. In this application, the effective magnetic moment was adjusted to approximately 1.4 by reducing the valence of $Li_2ScMo_3O_8$, as depicted in Figure S2. Consequently, it is evident that $Mo_3O_8$-type CMI, exhibiting an effective magnetic moment of approximately 1.4, demonstrates a temperature-dependent magnetic susceptibility amenable to the application of phenomenological the PCO model.

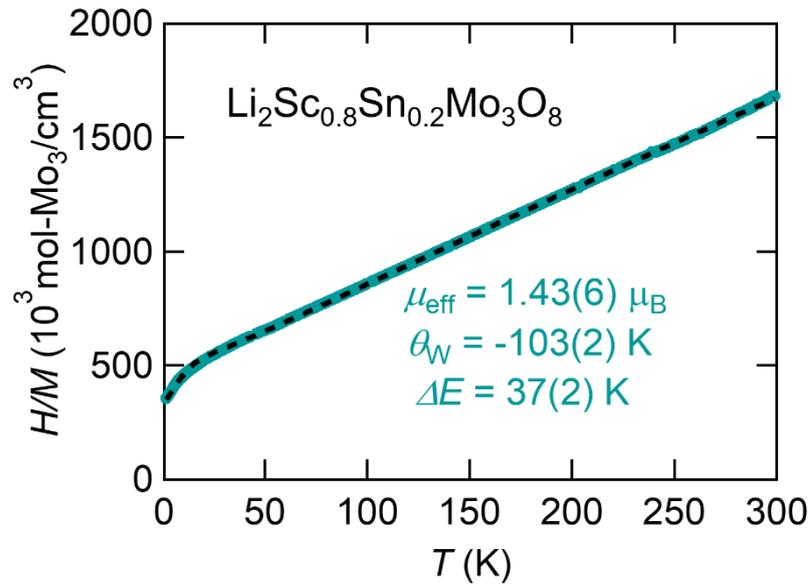

**Figure S2** The inversed magnetic susceptibility of $Li_2Sc_{0.8}Sn_{0.2}Mo_3O_8$. The dashed line shows the phenomenological PCO fit.